\documentclass[twocolumn,showpacs,preprintnumbers,amsmath,amssymb,prb,
superscriptaddress]{revtex4-1}

\usepackage{graphicx}
\usepackage{dcolumn}
\usepackage{bm}
\usepackage{dcolumn}
\usepackage{amsmath}
\usepackage{amssymb}
\usepackage{longtable}
\usepackage{array}
\usepackage[version=3]{mhchem}
\usepackage{epic}
\usepackage{epstopdf}
\usepackage{verbatim} 

\usepackage[usenames,dvipsnames,svgnames,table]{xcolor}

\newcolumntype{C}[1]{>{\centering}m{#1}}
\newcommand{\etal}{\textit{et al.}}
\newcommand{\bv}[1]{{\boldsymbol #1}}
\newcommand{\vsp}{\\[0.35cm]}
\newcommand{\muB}{\mu _\text{B}}

\begin{document}


\title{Theoretical investigation of the electronic and magnetic properties of the orthorhombic phase of \ce{Ba(Fe_{1-$x$}Co_$x$)_2As2}} 
\author{Gerald Derondeau}    \email{gerald.derondeau@cup.uni-muenchen.de}
\affiliation{%
  Department  Chemie,  Physikalische  Chemie,  Universit\"at  M\"unchen,
  Butenandstr.  5-13, 81377 M\"unchen, Germany\\}
\author{Svitlana Polesya} \affiliation{%
  Department  Chemie,  Physikalische  Chemie,  Universit\"at  M\"unchen,
  Butenandstr. 5-13, 81377  M\"unchen, Germany\\}
\author{Sergiy Mankovsky} \affiliation{%
  Department  Chemie,  Physikalische  Chemie,  Universit\"at  M\"unchen,
  Butenandstr. 5-13, 81377  M\"unchen, Germany\\}
\author{Hubert Ebert}
\affiliation{%
  Department  Chemie,  Physikalische  Chemie,  Universit\"at  M\"unchen,
  Butenandstr. 5-13, 81377 M\"unchen, Germany\\}
\author{J\'an Min\'ar}
\affiliation{%
  Department  Chemie,  Physikalische  Chemie,  Universit\"at  M\"unchen,
  Butenandstr. 5-13, 81377 M\"unchen, Germany\\}
\affiliation{%
  NewTechnologies-Research Center, University of West Bohemia, Pilsen, Czech Republic\\}

\date{\today}

\begin{abstract}
We present a comprehensive study on the low-temperature orthorhombic phase of \ce{Ba(Fe_{1-$x$}Co_$x$)_2As2} based on the Korringa-Kohn-Rostoker-Green function approach. Using this bandstructure method in combination with the coherent potential approximation alloy theory we are able to investigate the evolution of the magnetic and electronic properties of this prototype iron pnictide for arbitrary concentrations $x$, while dealing with the chemical disorder without uncontrolled simplifications by using solely a rigid band shift or the virtual crystal approximation. We discuss the development of the site resolved magnetic moments for the experimentally observed stripe antiferromagnetic order together with the strong electronic anisotropy of the Fermi surface and compare it with angle-resolved photoemission spectroscopy measurements of detwinned crystals. We furthermore calculate magnetic exchange coupling parameters $J_{ij}$ and use them for Monte-Carlo simulations on the basis of the classical Heisenberg model to get an insight on the temperature dependence of the magnetic ordering on the cobalt concentration.
\end{abstract}
                              
                             
\maketitle

%
\section{Introduction}
Over the last few years the iron pnictides received tremendous interest, following the discovery of high-temperature superconductivity in \ce{La(O_{1-$x$}F_$x$)FeAs}.\cite{KWHH08,TIA+08} Its mechanism of superconductivity is generally considered to be unconventional and it is most likely connected to magnetic fluctuations.\cite{Sin08a,MSJD08,FPT+10} This makes the magnetic behavior of the iron pnictides crucial to understand their underlying physics and superconductivity. However, this question turned out to be far from trivial.\cite{MJ09, MJB+08} The complex magnetism of these compounds allows no straightforward description concerning several aspects which results in the fact that even today the iron pnictides are far from fully understood.

To name a few examples, there was considerable discussion whether the magnetic moments are better described by an itinerant\cite{MJ09,MSJD08,YLAA09,OKZ+09,FTO+09} or a localized\cite{HYPS09,YZO+09,CLLR08} model, there is still no consensus over the strength of correlation effects\cite{WCM+12,LYM+08} and finally the magnitude of the magnetic moments is highly sensible on the system and computational parameters, which leads to several seemingly quite different reports in literature.\cite{MJB+08,SLS+09,RTJ+08,GLK+11,VFB+12} Experimental neutron diffraction data predicts for the low-temperature phase of \ce{BaFe2As2} magnetic moments around $0.9\muB$ per Fe atom ($0.99\muB$ in Sn flux\cite{SLS+09}, $0.87\muB$ for powder probes\cite{HQB+08}), while from \ce{^{57}Fe} M\"ossbauer spectroscopy\cite{RTJ+08,RTS+09} and $\mu$SR spectroscopy\cite{ABB+08,GAB+09} consistently a value of around $0.5\muB$ is estimated. However, in density functional theory (DFT) calculations the magnitude of the magnetic moments is considerably overestimated, ranging from approximately $1.2\muB$ up to $2.6\muB$.\cite{KOK+09,YLAA09,SBP+11,MJB+08,AC09} Furthermore, it is well known that the magnetic moment depends surprisingly strong on the free structural parameter $z$ of the As position, introducing another degree of freedom which makes reliable predictions even more difficult.\cite{MJB+08,YLH+08,HYPS09}

It is believed that the commensurate magnetic spin-density-wave (SDW) state and the superconducting state compete with each other, implying that the suppression of the long range magnetic order is coupled to the emergence of superconductivity.\cite{Sin08a,PKK+11,RTS+09} Consequently, the understanding of the magnetic state is crucial to understand the superconducting behavior of these compounds.

In this paper we address the magnetic state of the undoped mother compound \ce{BaFe2As2} and focus on the impact of chemical disorder effects induced by substitution of Fe by Co. Most theoretical studies on the doping dependence of iron pnictides are based on a virtual crystal approximation (VCA) which introduces an averaged atomic charge $Z$ for atomic sites with chemical disorder.\cite{KOK+09,Sin08a,PIY+10} Using the VCA one should keep in mind that site resolved information is lost and disorder is not properly described. Recent publications stressed that it is not sufficient for the iron pnictides to neglect these more complex disorder effects and proved the necessity of more sophisticated approaches.\cite{BLGK12, WBW+13} As an example, impurity scattering is discussed to be the crucial aspect for the newly discovered in-plane resistivity anomaly in \ce{Ba(Fe_{1-$x$}Co_$x$)_2As2} which shows the pressing need to account for disorder effects in an appropriate way within a theoretical description.\cite{CAG+10, INL+13, GHA14}

It is very difficult for wave-function based methods to achieve a reasonable inclusion of disorder effects, the most common way is to use supercells.\cite{BLGK12, LSC+12} The major disadvantage of such an approach is the high computational effort which limits its possible applications. Using a Korringa-Kohn-Rostoker-Green function (KKR-GF) based method the coherent potential approximation (CPA) is a more straightforward way to account for disorder compared to a supercell calculation but with considerably less computational effort needed. Up to now, investigations of the iron pnictides using the CPA are extremely rare and recent in literature but nevertheless very promising.\cite{KJ14, KAJ14} In this paper we will exploit the significant advantages of the CPA method to deal with the substitution induced disorder in iron pnictides to achieve an improved theoretical description for the doping dependent evolution of these compounds.  
%
\section{Computational approach}
\begin{figure}[tb]
\input{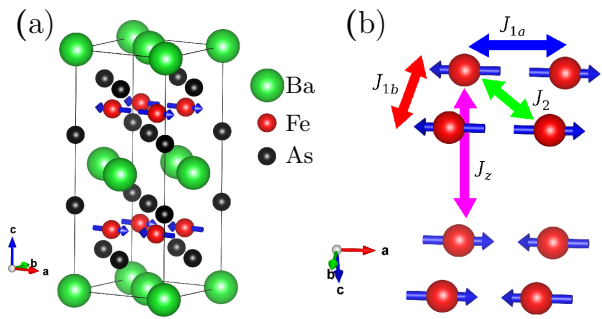}
   \caption{(Color online) Crystal and magnetic structure of orthorhombic \ce{BaFe2As2}. The blue arrows on Fe indicate the spin magnetic moments. (a) Conventional orthorhombic $Fmmm$ low-temperature unit cell. (b) Magnetic structure of the Fe atoms with the nearest- and next-nearest-neighbor exchange interactions. The color code of the $J_{ij}$ values corresponds to Fig.~\ref{Fig_JXC1}.}\label{Fig_Crystal}
\end{figure}

All calculations have been performed self-consistently and fully relativistically within the four component Dirac formalism, using the Munich SPR-KKR program package\cite{EKM11}. We used always the LDA exchange-correlation potential with the parameterization given by Vosko, Wilk and Nusair.\cite{VWN80} The structural setup was based on an orthorhombic unit cell of \ce{BaFe2As2} with four Fe atoms per cell (4-Fe unit cell) in order to account for the experimentally observed stripe antiferromagnetic spin state with antiferromagnetic coupling along $a$ and $c$ and ferromagnetic chains along $b$ (see Fig.~\ref{Fig_Crystal}a). With spin-orbit coupling included by the four component Dirac formalism the self-consistent field (SCF) calculations considered an orientation of the magnetic moments along the $a$ axis, consistently with experiment.\cite{SLS+09} We used a dense $\bv k$-mesh of $20 \times 18 \times 20$ points and considered for $s$, $p$ and $d$ orbitals as basis. As spherical approximation we used a so-called full-charge ansatz which uses Voronoi polyeder as within the full-potential scheme. Although all aspherical parts of the charge density are fully accounted for, the aspherical parts of the potential are neglected in a full-charge ansatz. We confirmed that the electronic structure in a full-charge calculation is comparable to the results from a real full-potential calculation but is achievable with strongly reduced computational effort. The exchange coupling constants $J_{ij}$ were calculated using the Lichtenstein formula.\cite{LKAG87,MBM+09,EM09a} The definition of the various $J_{ij}$ Fe-Fe coupling parameters between Fe atoms is shown in Fig.~\ref{Fig_Crystal}b. The treatment of disorder introduced through Co substitution is fully dealt with on a CPA level.

Within this work we used the experimental lattice parameters and the experimental As position $z$ of \ce{BaFe2As2}.\cite{RTJ+08} However, in case of a large concentration regime ($0 \leq x_\text{Co} \leq 0.25$) it is reasonable to consider structural relaxation. Thus, we used the experimentally observed lattice constants from Sefat \etal \cite{SJM+08} for \ce{BaFe2As2} and \ce{Ba(Fe_{0.9}Co_{0.1})2As2} and extrapolated on this basis the change in the crystallographic $c$-parameter under Co substitution in the orthorhombic phase. The lattice constants $a$ and $b$ were not changed for the calculations, because their deviation in experiment is reported small enough to be assumed as unchanged within experimental uncertainty.\cite{SJM+08}  The validity of such an  extrapolation is further supported by other work which shows similar trends in the lattice parameters.\cite{NTY+08, KOK+09} The change of the As position was accounted for on the basis of a publication by Merz \etal \cite{MSN+13}, where we also considered for the clinching of $c$ for higher doping values to get the best possible extrapolation. All used structure parameters are summarized in Tab.~\ref{Tab_Lattice} of the appendix.

The phase transition from orthorhombic to tetragonal was intentionally not considered, because we wanted to focus on the magnetic state and have the results comparable over the whole doping regime.
%
\section{Results and Discussion}
\subsection{Magnetic Moments}
For the undoped orthorhombic mother-compound \ce{BaFe2As2} we found a total magnetic moment of $1.19\muB$, having a spin magnetic moment of $1.14\muB$ and an orbital magnetic moment of $0.05\muB$. It should be noted that this is in quite reasonable agreement with experimental neutron diffraction data\cite{SLS+09, HQB+08} compared to other literature, considering that we used a LDA exchange-correlation potential and the experimental As position without structural optimization.\cite{SBP+11,MJB+08,MJ09,AC09}
\begin{figure}[tb]
\vspace{-0.5cm}
\hspace{+1.5cm}\scalebox{0.70}{\input{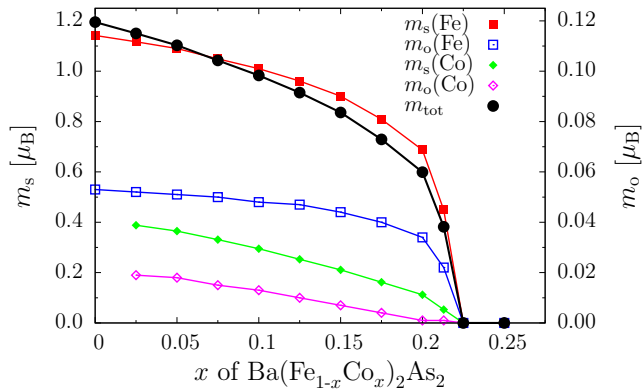}}
\vspace{-0.8cm}
\caption{(Color online) Magnetic spin moments ($m_\text{s}$) and orbital moments ($m_\text{o}$) of Fe and Co for increasing Co ratio $x$. The black line corresponds to the total magnetic moment which is the substitution dependent sum over all other plotted contributions. To show spin and orbital moments in one graph the black, red and green curve correspond to the left $y$-axis while the blue and magenta plot belong to the right $y$-axis which is reduced by one order of magnitude.}\label{Fig_MagnMom}
\end{figure}

In Fig.~\ref{Fig_MagnMom} we show the evolution of the different contributions to the total magnetic moment depending on the increasing substitution of Co on the Fe sites. For each doping ratio $x$ the system was calculated fully self-consistently with the CPA. The use of VCA calculations is insufficient for such an investigation, not only because of the intrinsic deficiencies of the made assumptions, but also because the VCA looses every site resolved information. Here we can distinguish the magnetic spin and orbital moments of Fe and Co respectively. Considering the lowest investigated Co concentration of 2.5~\% per Fe in \ce{Ba(Fe_{0.975}Co_{0.025})2As2} one has a total magnetic moment of $1.15\muB$ with already a noteworthy reduction compared to the undoped case ($1.19\muB$). On the one hand side, this reduction is approximately due to the individual decrease of the Fe magnetic moments and on the other hand due to the smaller contribution of Co. In fact, the magnetic moments of Co are almost by a factor of 3 smaller compared to the Fe magnetic moments. This is a strong difference compared to the bulk metals bcc-Fe and hcp-Co where the difference in the magnetic moments is small and only about 36~\%. On the other hand, \ce{BaFe2As2} has a clear magnetic transition with finite moments while \ce{BaCo2As2} is a solely paramagnetic metal without any signs for a magnetic transition.\cite{SSJ+09}

For increasing doping ratio $x$ obviously the influence of Co on the decrease in the magnetic moments gets more and more pronounced. Considering only the Co magnetic spin and orbital moments it is obvious that they decrease more or less linearly until they vanish completely. The decrease in the Fe magnetic moments is less pronounced for low doping concentrations but gets significantly higher for $x$ values above 0.1. In the $x$ range between 0.2 and 0.225 the Co moments finally vanish, which is accompanied by a drastic collapse of the Fe magnetic moments and the system becomes paramagnetic. This collapse of antiferromagnetic order differs from the experimental phase diagram\cite{LCA+09} of \ce{Ba(Fe_{1-$x$}Co_$x$)_2As2} because the experimentally observed disappearance of long range magnetic order is coupled to an incommensurate SDW state or fluctuating magnetic moments and not to a non-magnetic state. 

Nevertheless, it gives a very interesting insight on the influence of Co doping on the magnetic order in \ce{BaFe2As2} and shows a clear non-linear behavior. It further quantifies how increasing substitution of Co on Fe sites weakens the fixed commensurate SDW state until the influence becomes strong enough to totally suppress magnetic long range order. Although this gives no direct information about the possible existence of another spin state with fluctuating magnetic moments, Co substitution proves to be quite efficient in suppressing the commensurate SDW state in \ce{BaFe2As2} which is obviously its crucial influence. 

%
\subsection{Bloch spectral functions}
\begin{figure}[tb]
\input{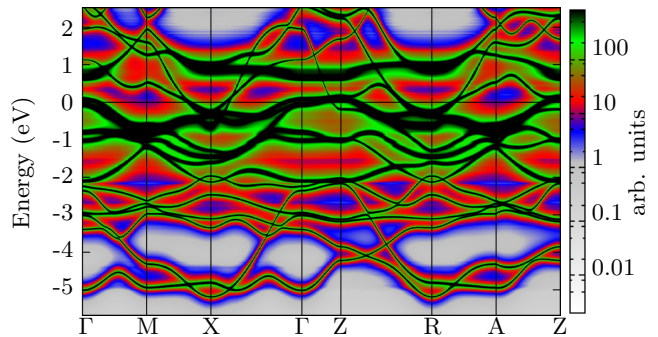}
\caption{(Color online) Bloch spectral function of non-magnetic, tetragonal \ce{Ba(Fe_{0.875}Co_{0.125})_2As2} calculated with the CPA, using one cell with five atoms. The corresponding Brillouin zone can be found in Fig.~\ref{Fig_BZTet} in the appendix.}\label{Fig_CPA}
\end{figure}

First we demonstrate that the CPA is able to reproduce all disorder effects necessary for a complete description of the electronic structure. Berlijn \etal \cite{BLGK12} calculated the bandstructure of non-magnetic, tetragonal \ce{Ba(Fe_{0.875}Co_{0.125})_2As2} for 10 randomly configured supercells containing 400 atoms each in order to gain a description of disorder effects. In Fig.~\ref{Fig_CPA} we show the Bloch spectral function for the same system calculated with the CPA but using only one cell having five atoms. All important aspects, explicitly the disorder induced band broadening, are in line with Berlijn's results\cite{BLGK12} concerning intensity as well as position in the reciprocal space. Thus, we can assume applicability of the CPA to treat the disorder in doped iron pnictides.

To further investigate the anisotropic effects of the stripe antiferromagnetic order depending on the Co concentration we compare the shapes of the different Fermi surfaces (FS) with each other. To evaluate the electronic structure we used the converged potentials to calculate corresponding Bloch spectral functions at the Fermi level which should reveal the strong in-plane anisotropy. To see these important anisotropic effects in angle resolved photoemission spectroscopy (ARPES) experiments it is necessary to efficiently detwin the single crystal after the structural and magnetic phase transitions at low temperatures, for example by applying in-plane uniaxial stress. Corresponding precise ARPES measurements of detwinned crystals of \ce{BaFe2As2} which show the strong in-plane anisotropy of the electronic structure are indeed available in literature.\cite{YLC+11, KOK+11}
\begin{figure}[tb]\centering 
\input{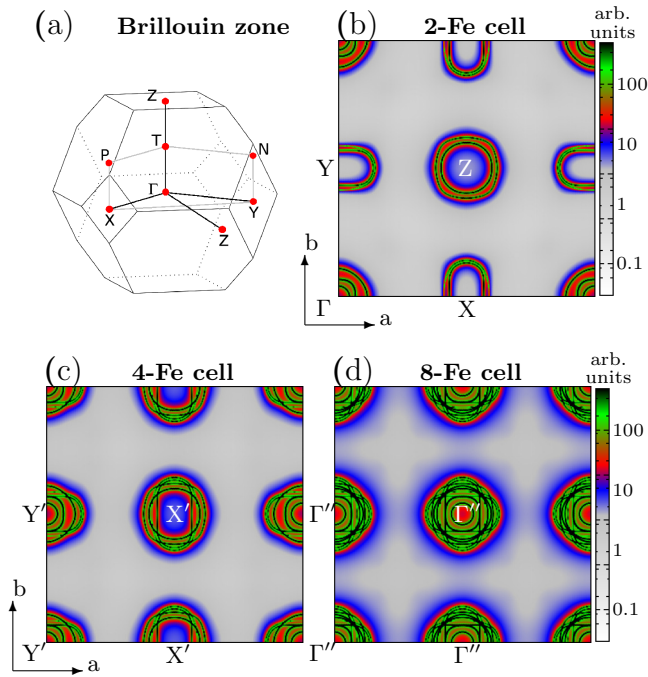}
\caption{(Color online) (a) Brillouin zone of orthorhombic \ce{BaFe2As2}. Corresponding Fermi surfaces are derived from non-magnetic calculations in the $\Gamma$XY plane for increasing size of the unit cell, (b) two Fe per cell (c) four Fe per cell (d) eight Fe per cell, showing increasing back-folding of the bands. 
}\label{Fig_FS_NonMag}
\end{figure}

Concerning calculations one should keep in mind that the FS depends strongly on the magnetic moment which is not easy to capture adequately. Furthermore one must understand the  increasing back-folding of bands with simultaneously increasing number Fe atoms per unit cell. Still, to describe the experimentally observed stripe antiferromagnetic state with antiferromagnetic coupling along $c$ one needs to consider at least a unit cell with four different Fe atoms (4-Fe). The effects of back-folding are schematically explained in Fig.~\ref{Fig_FS_NonMag} where we show the corresponding Brillouin zone (BZ) (Fig.~\ref{Fig_FS_NonMag}a) and compare the Fermi surfaces of orthorhombic undoped \ce{BaFe2As2} derived from strict non-magnetic calculations of a 2-Fe (Fig.~\ref{Fig_FS_NonMag}b), a 4-Fe (Fig.~\ref{Fig_FS_NonMag}c) and a 8-Fe (Fig.~\ref{Fig_FS_NonMag}d) unit cell. In the primitive 2-Fe unit cell (Fig.~\ref{Fig_FS_NonMag}b) all points $\Gamma$, $\mathrm X$, $\mathrm Y$ and $\mathrm Z$ are distinct. Note, that the directions of $\mathrm X$ and $\mathrm Y$ correspond directly to the real space $a$ and $b$ directions. In the 4-Fe unit cell (Fig.~\ref{Fig_FS_NonMag}c) the back-folding results in two points we will call $\mathrm Y'$ and $\mathrm X'$ where $\mathrm Y'$ is a superposition of $\Gamma$ and $\mathrm Y$ while $\mathrm X'$ is the corresponding superposition of $\mathrm X$ and $\mathrm Z$. Further back-folding results in Fig.~\ref{Fig_FS_NonMag}d for the case of a 8-Fe unit cell where only one point $\Gamma ''$ is the superposition of $\Gamma$, $\mathrm X$, $\mathrm Y$ and $\mathrm Z$. Because the calculations of the stripe magnetic state require at least the 4-Fe unit cell the Fermi surfaces will correspond to an equivalent back-folding with only $\mathrm X'$ and $\mathrm Y'$ being distinct.

In Fig.~\ref{Fig_FS1}a we show the Fermi surface of undoped \ce{BaFe2As2} in its stripe antiferromangetic state. Even considering the expected 4-Fe unit cell back-folding the strong in-plane anisotropy of the electronic structure is obvious. The red points correspond to the reconstructed Brillouin zone of Yin \etal \cite{YLC+11} which was derived from ARPES measurements on detwinned crystals. This reconstructed BZ is a combination of all measured ARPES data and would hence correspond to an overlay of $\mathrm X'$ and $\mathrm Y'$ in the definition of Fig.~\ref{Fig_FS_NonMag}c. The good agreement with the experimental ARPES data is obvious. We loose in the calculations the inner circles of the reconstructed BZ which would arise from bands around $\Gamma$ but the most important bands, namely the anisotropic ones are strikingly well preserved. They are described in literature as small bright spots along $\mathrm X$ (corresponding to the antiferromagnetic real-space $a$ direction) and larger petals along $\mathrm Y$ (corresponding to ferromagnetic real space $b$ direction).\cite{YLC+11} The bright spots along $\mathrm X$ are perfectly reproduced. The petals along $\mathrm Y$ are a bit bigger than in experiment, but their characteristics are clearly identifiable. More or less comparable Fermi surfaces for the stripe antiferromagnetic iron pnictides were predicted for example by Andersen and Boeri.\cite{AB11} 

\begin{figure}[tb]
\input{pictex/FS1.tex}
\caption{(Color online) Fermi surfaces of the experimental stripe antiferromagnetic state of orthorhombic \ce{Ba(Fe_{1-$x$}Co_$x$)_2As2} calculated with a 4-Fe unit cell for different Co concentrations $x$. (a) undoped \ce{BaFe2As2} with an overlay of the reconstructed Brillouin zone of Yin \etal \cite{YLC+11} from ARPES measurements (b) \ce{Ba(Fe_{0.975}Co_{0.025})_2As2} (c) \ce{Ba(Fe_{0.925}Co_{0.075})_2As2} (d) \ce{Ba(Fe_{0.875}Co_{0.125})_2As2}.}\label{Fig_FS1}
\end{figure}
\begin{figure}[htb]
\input{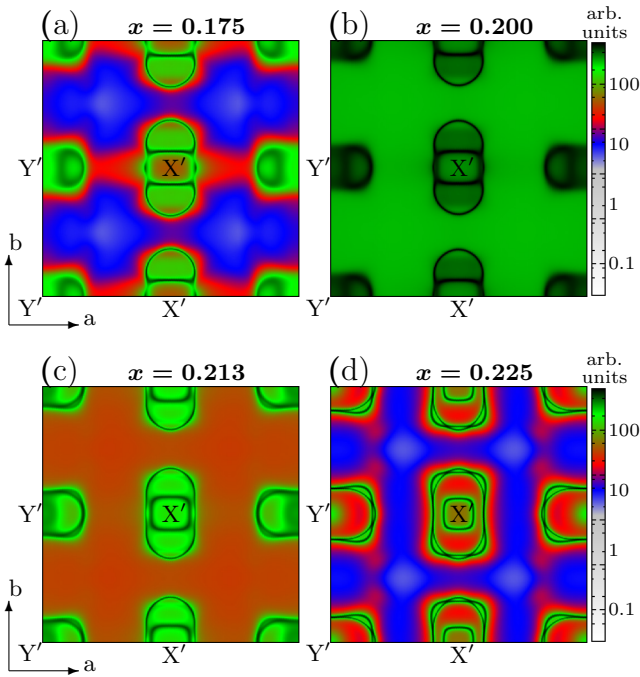}
\caption{(Color online) Fermi surfaces of the experimental stripe antiferromagnetic state of orthorhombic \ce{Ba(Fe_{1-$x$}Co_$x$)_2As2} calculated with a 4-Fe unit cell for different Co concentrations $x$. These Fermi surfaces are close to the collapse of long range antiferromagnetic order, (a) \ce{Ba(Fe_{0.825}Co_{0.175})_2As2} (b) \ce{Ba(Fe_{0.8}Co_{0.2})_2As2} (c) \ce{Ba(Fe_{0.787}Co_{0.213})_2As2} (d) \ce{Ba(Fe_{0.775}Co_{0.225})_2As2}.}\label{Fig_FS2}
\end{figure}

In addition to the undoped Ba-122 compound we performed the calculation of Bloch spectral functions for the whole Co doping regime considered in this work. The CPA allows a very precise investigation of the influence of Co on the electronic structure considering the induced chemical disorder. Furthermore it is interesting to see the change in the Fermi surface for decreasing strength of the long-range antiferromagnetic order. In Fig.~\ref{Fig_FS1}b the shape of the FS has not drastically changed, however a blur in the intensity is already quite visible. These blurs or band broadening effects are due to the disorder induced by Co. For the higher concentrations in Fig.~\ref{Fig_FS1}c and Fig.~\ref{Fig_FS1}d the shape of the bright spots along $a$ starts to change. While the spots around $\mathrm X'$ shrink the ones around $\mathrm Y'$ increase in size and start to blur out. This trend continues as can be seen in Fig.~\ref{Fig_FS2}a which shows for $x=0.175$ still a comparable FS before the antiferromagnetic order collapses for higher Co concentrations. The initially bright spots around $\mathrm Y'$ have developed to petal like structure like the ones around $\mathrm X'$ along the $b$ direction. However, they are still clearly distinguishable through the different strength of band blurring. The crucial change of the FS happens in Fig.~\ref{Fig_FS2}b where the anisotropic features finally start to vanish and become symmetric propeller-like structures. Note however, that the electronic structure along $b$ direction, with ferromagnetic coupling, is perfectly sharp with no signs of disorder effects, while the band blur along the $a$ direction, with antiferromagnetic coupling, is strongest in this picture. For $x=0.213$ in Fig.~\ref{Fig_FS2}c the matching of the propeller structures is already nearly perfect although there is still a finite magnetic moment on Fe. Additionally, a square like feature forms around $\mathrm X'$ which originates most likely through the back-folding from the original $\mathrm Z$ point in the 2-Fe unit cell. The collapse of long range magnetic order is complete in Fig.~\ref{Fig_FS2}d where now the propeller structures are perfectly symmetric. There are clearly no in-plane anisotropic features of the electronic structure left. Note that the absence of a perfect fourfold rotational symmetry of the FS is only due to the effects of back-folding in the $\Gamma$XY plane (compare Fig~\ref{Fig_FS_NonMag}). 

\begin{figure}[tb]
\input{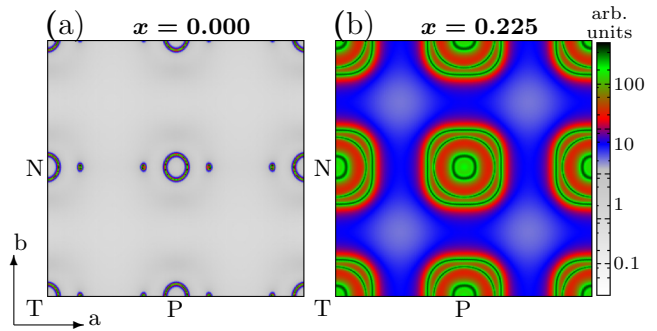}
\caption{(Color online) Fermi surfaces of \ce{Ba(Fe_{1-$x$}Co_$x$)_2As2} calculated with a 4-Fe unit cell in the TPN plane, which is shifted by $\bv k_\text z = 0.5$ in $\mathrm Z$ direction. (a) For $x=0.0$ the fourfold rotational symmetry in the magnetic state is clearly broken. (b) For $x=0.225$ in the paramagnetic state the fourfold rotational symmetry is in principal perfectly restored, despite the 4-Fe unit cell and despite the orthorhombic distortion.}\label{Fig_FS3tnp}
\end{figure}

The same effect is more clearly visible if one shifts the Bloch spectral function from the $\Gamma$XY plane with $\bv k_\text z = 0.5$ towards the $\mathrm Z$ point. The corresponding TPN plane has the property to show the restoration of the fourfold rotational symmetry after the collapse of long-range magnetic order despite the back-folding of bands in a 4-Fe unit cell. While Fig.~\ref{Fig_FS3tnp}a for the magnetic state with $x=0.0$ has an obvious twofold rotational symmetry, the fourfold rotational symmetry is clearly restored after the collapse of the long-range antiferromagnetic order in Fig.~\ref{Fig_FS3tnp}b for $x=0.225$. Note further that this disappearance of the in-plane anisotropy occurs for an orthorhombic lattice. Consequently, one can state that the origin of the strong anisotropy in ARPES measurements is practically only due to the stripe antiferromagnetic order while the effect of the lattice distortion is practically neglectable.

It is also striking that the band blurring effects for $x=0.225$ (see Fig.~\ref{Fig_FS2}d) are significantly reduced compared to for example $x=0.2$ (see Fig.~\ref{Fig_FS2}b). This is surprising as one would expect a continuous increase of the band blurring effects to the maximum of substitutional disorder at 50~\% Co substitution. The deviation from the expected behavior is obviously connected to the antiferromagnetic order and its collapse for  $x=0.225$. Additionally, the in-plane anisotropy of the disorder in Fig.~\ref{Fig_FS2}b may be an interesting first indication for anisotropic effects in the transport properties at the proximity of the magnetic phase. It has recently been discussed that these effects are possibly due to anisotropic scattering properties.\cite{INL+13} Still, an investigation of only the Fermi surface is insufficient for even qualitative statements and specific calculations of the linear response properties\cite{KLSE11} of these systems are necessary and planed for the future.
%
\subsection{Exchange Coupling Constants}
To further investigate the magnetic structure of orthorhombic \ce{Ba(Fe_{1-$x$}Co_$x$)_2As2} we investigated the exchange-coupling constants $J_{ij}$ between sites $i$ and $j$. Calculations are based on the magnetic force theorem and implemented in the multiple scattering formalism by the formula of Lichtenstein \etal\cite{LKAG87,MBM+09,EM09a}.

We focus our discussion on the most important coupling constants $J_{1a}$, $J_{1b}$, $J_2$ and $J_z$ between neighboring Fe atoms, defined according to Fig.~\ref{Fig_Crystal}b. Experimentally, it is often not possible to directly determine these values independently but one fits the experimental data based on a chosen model, e.g. a Heisenberg model.\cite{EPB+08,ZYL+08} Sometimes the relative strength of the different coupling constants is further fixed in their ratio based on theoretical calculations while fitting.\cite{EPB+08} Although there are numerous calculations in literature on the relative and absolute strength of the coupling constants there is no clear consensus and the published values differ in magnitude and ratio.\cite{EPB+08,ZYL+08, YLH+08,HYPS09, Yil08,MLX08} Thus, it is difficult to compare the results due to different definitions of the exchange coupling constants, the different approximations to calculate them and because the calculated magnetic moments may differ significantly. For the iron pnictides it is known that the exchange energies are quite sensitive with respect to the magnitude of the magnetic moment, which consequently may result in different absolute values.\cite{YLH+08,HYPS09,Yil08}

The exchange coupling constants discussed here refer to a Fe atom in center coupling with the neighboring Fe atoms or substituted Co atoms on Fe sites. The corresponding classical Heisenberg Hamiltonian has the form $\mathcal{H} = - \sum_{ij} J_{ij} \mathbf{e}_i \mathbf{e}_j$, where we use unit vectors $\mathbf{e}_{i(j)}$ instead of spin vectors $\mathbf{S}_{i(j)}$.\cite{MBM+09,PKM+12} Here, a negative sign corresponds to antiferromagnetic coupling while a positive sign favors a ferromagnetic interaction. The coupling constants between Fe and Co behave in principal similarly, only the absolute values of Co are reduced by a factor of approximately 3, which could be expected as it is a similar ratio compared to the relative strength of the magnetic moments.
\begin{figure}[tb]
 \scalebox{0.68}{\input{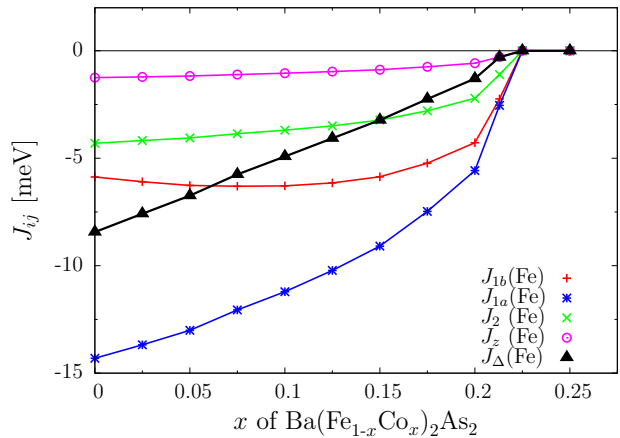}}
\caption{(Color online) Isotropic exchange coupling constants $J_{ij}$ of Fe coupling with neighboring Fe atoms. $J_\Delta$ is defined as $J_{1a} - J_{1b}$.}\label{Fig_JXC1}
\end{figure}

We plotted the isotropic exchange coupling constants for Fe coupling with Fe as a function of the Co doping ratio $x$ in Fig.~\ref{Fig_JXC1}. Here $J_\Delta$ is defined as $J_{1a} - J_{1b}$ and corresponds to the real-space in-plane anisotropy between $a$ and $b$. As can be seen in the plot $J_\Delta$ decreases nearly perfectly linear for increasing Co doping. This smooth diminishing in-plane anisotropy of \ce{BaFe2As2} is again induced by the substitution of Co. $J_{1a}$ and $J_{1b}$ are the dominating interaction parameters with strong dependence on the Co doping, while $J_{2}$ and $J_{z}$ are smaller and their change due to Co substitution is much less significant. 
The derived coupling constants are in reasonable agreement with experimental neutron diffraction data on \ce{BaFe2As2}.\cite{EPB+08,QSW+12,HLL+11} The values seem a bit smaller but this is primarily due to the definition of the Heisenberg Hamiltonian (see above) with the spin moment incorporated into the $J_{ij}$'s.\cite{MBM+09,PKM+12}

It should be noted, that the signs for all calculated nearest neighbor interactions are negative and would hence prefer an antiferromagnetic order. As the system can obviously not fulfill all of these conditions at the same time there is a competition of magnetic states, in accordance with other reports in literature.\cite{Yil08,SA08} 
%
%
\subsection{Monte Carlo simulations}
To solve the problem of competing magnetic states on an accurate level we performed Monte Carlo simulations based on the classical  Heisenberg model. It should be noted, that although the iron pnictides have clearly itinerant aspects in their magnetic structure it is still too rash to completely dismiss the Heisenberg model for predictions of the magnetic ordering. On a simple level the Heisenberg model was successfully used in several publications giving useful results as long as one keeps the underlying approximations in mind.\cite{ZYL+08, EPB+08, Yil08}

\begin{figure}[tb]
\setlength{\unitlength}{1.0cm}
\begin{picture}(12,6)
\put(0.2,5.5){\large \textbf (a)}
\put(3.5,5.5){\large \textbf (b)}
\put(0.0,0.7){\includegraphics[width=3.9\unitlength,clip]{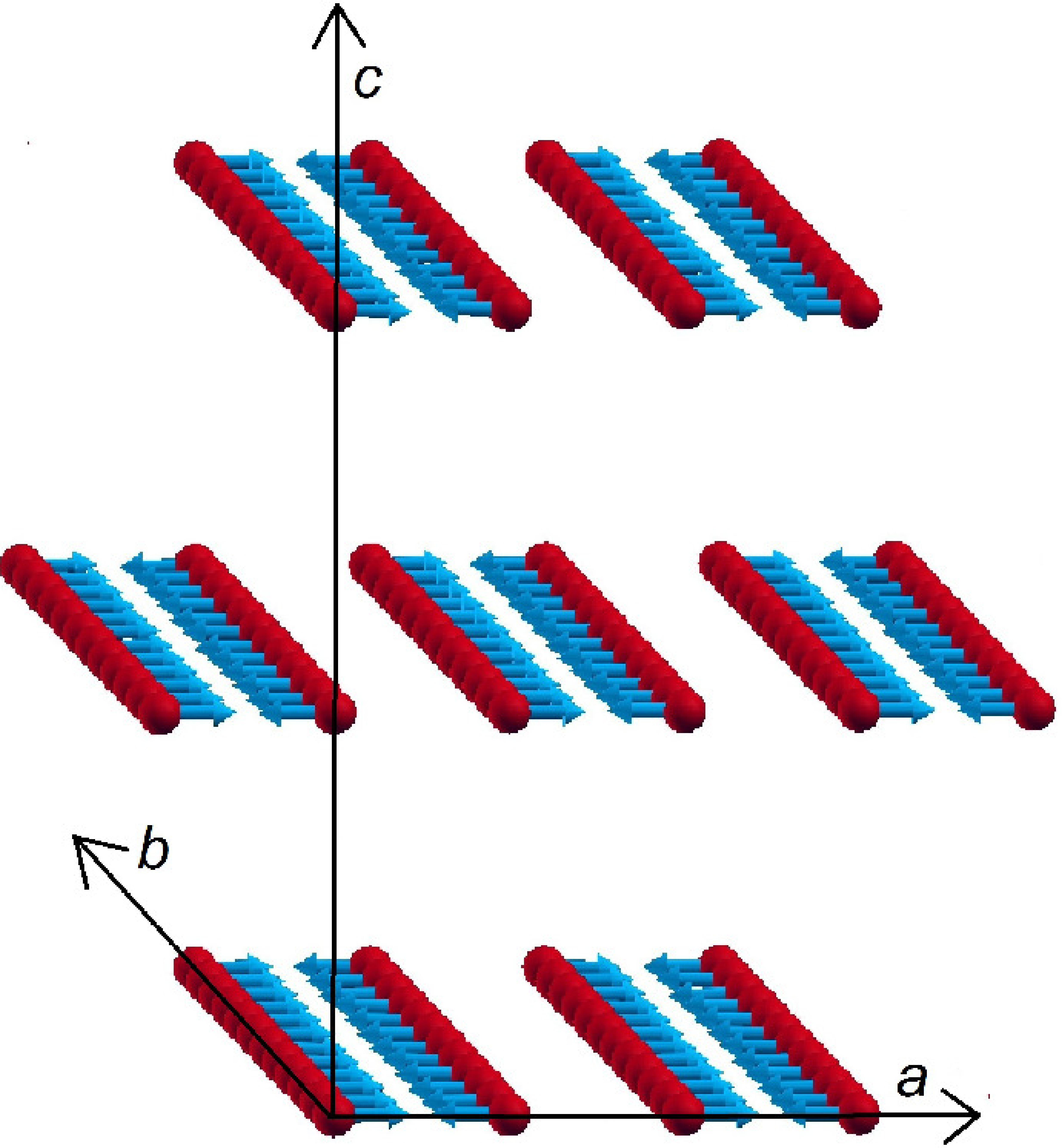}}
\put(1.8,0.0){\scalebox{0.68}{\input{pictex/TNeel.tex}}}
\end{picture}
\caption{(Color online) (a) Magnetic ground state of \ce{BaFe2As2} according to Monte Carlo simulations. (b) Evolution of the N\'eel temperatures $T_\text{N}$ depending on the Co concentration $x$ in \ce{Ba(Fe_{1-$x$}Co_$x$)_2As2}. The experimental data $T_\text{N, exp}$ was reproduced from the work of Lester \etal \cite{LCA+09}.}\label{Fig_TNeel}
\end{figure}

For the Monte Carlo simulations we used 2744 atoms and solved the competition of magnetic states in an accurate way as we reproduced correctly the experimental stripe antiferromagnetic structure as magnetic ground state for T~=~0~K as seen in Fig.~\ref{Fig_TNeel}a. These calculations allowed us further to include temperature dependent effects to evaluate for example the N\'eel temperature $T_\text N$ of the ground state. For the undoped \ce{BaFe2As2} the estimated N\'eel temperature of $T_\text N \approx 142~\text{K}$ is in perfect agreement with the experimental  $T_\text{N, exp} \approx 140~\text{K}$. These results make us confident that the exchange coupling constants are in the right order of magnitude and the Heisenberg model can be successfully used as a satisfying basis for Monte Carlo simulations.

We further evaluated the change of the N\'eel temperature for increasing Co substitution and show this behavior in Fig.~\ref{Fig_TNeel}b together with experimental data taken from the work of Lester \etal \cite{LCA+09}. One should still keep in mind that the calculated collapse of antiferromagnetic order happens for relatively high Co ratios, meaning it cannot be directly compared to the experimental phase diagram concerning the Co concentration as we did not consider for incommensurate spin states. 

Still, accounting for this stretching of the calculated phase diagram the overall trends in the behavior of the N\'eel temperature in Fig.~\ref{Fig_TNeel}b are rather well reproduced. Also in the experimental data from Lester \etal \cite{LCA+09} one can see a split of the Co dependence of the N\'eel temperature in two approximately linear regions with different slope. Experimentally, in the first region $T_\text{N}$ decreases from 140~K down to approximately 40~K until the proximity of the magnetic collapse is reached. For higher Co ratios the decrease of $T_\text{N}$ from 40~K down to 0~K (collapse of long-range magnetic order) goes faster, leading to an almost perpendicular drop of the N\'eel temperature. In the calculated N\'eel temperatures one can see also these two regions with the same behavior, even for the same values of $T_\text{N}$.

%
\section{Summary}
In this work we evaluated the magnetic and electronic properties of \ce{Ba(Fe_{1-$x$}Co_$x$)_2As2} in its orthorhombic, stripe antiferromagnetic ordered state. The substitution of Co on Fe sites was dealt within the CPA which allows in a quite sophisticated way to deal with the disorder of the system on a level beyond a VCA or a rigid-band shift calculation. We further showed, that the CPA results are fully in line with expensive supercell calculations of Berlijn \etal \cite{BLGK12}.

We calculated the site resolved magnetic spin and orbital moments fully relativistically over a wide doping regime from $x=0.0$ to $x=0.25$ in steps of 2.5~\%. For the undoped \ce{BaFe2As2} we got a total magnetic moment of $1.19 \muB$ per Fe which is smaller and closer to experimental neutron data compared to other publications which used LDA and the experimental, not optimized As position $z$.\cite{SBP+11,MJB+08,MJ09,AC09} We found decreasing magnetic moments on Fe and Co in \ce{Ba(Fe_{1-$x$}Co_$x$)_2As2} for increasing Co concentration as expected until the system becomes paramagnetic at $x=0.225$. Although this collapse of long range magnetic order cannot be directly compared to the disappearance of antiferromagnetic order in the phase diagram, it is still important concerning the decreasing stability of a commensurate SDW state under Co substitution.

The change of the electronic structure at the Fermi level was also evaluated with special focus on the strong in-plane anisotropy observed in ARPES measurements. We were able to reproduce the anisotropic feature of the Fermi surface in an adequate way and observed that this anisotropy decreases together with the magnetic moments until it completely vanishes where the system becomes paramagnetic. As this was obtained in an orthorhombic lattice it is quite clear that the origin of this exceptional strong electronic anisotropy is only due to the stripe magnetic order while the lattice distortion effects between $a$ and $b$ can be neglected.

Furthermore, we calculated the isotropic exchange coupling constants $J_{ij}$ for Fe and Co. We observed strong in-plane anisotropy between $J_{1a}$ and $J_{1b}$ which again consequently decreases for increasing Co concentration. The coupling constants had the same sign for all nearest neighbor interactions, indicating competition of magnetic states and showed reasonable agreement with experimental values.\cite{EPB+08,QSW+12,HLL+11} The dilemma of competing magnetic states could be successfully solved by Monte Carlo simulations based on a classical Heisenberg model, which reproduced almost perfect N\'{e}el temperatures and the stripe antiferromagnetic spin state as magnetic ground state.

Overall, we showed successfully the decrease of the strong in-plane anisotropy \ce{Ba(Fe_{1-$x$}Co_$x$)_2As2} for increasing Co concentration within the CPA. Thus, the CPA has proven to be a highly valuable and precise tool to investigate the influence of chemical disorder introduced by substitution of elements which is crucial for understanding the iron pnictides. 
%
%
\section*{Acknowledgments}
We like to thank Alexander Yaresko for his extraordinary help and valuable discussions. Further thank goes to Dirk Johrendt for his help. We acknowledge the financial support from the  Deutsche Forschungsgemeinschaft DFG (projects FOR 1346 and EB 154/26-1) and from the Bundesministerium f\"ur Bildung und Forschung BMBF (project 05K13WMA). We further thank for the support from CENTEM (project CZ.1.05/2.1.00/03.0088).
%
%
\section*{Appendix}
\begin{figure}[htb]
\setlength{\unitlength}{1.0cm}
\begin{picture}(12,3.0)
\put(2.0,0.0){\includegraphics[width=3.5\unitlength,clip]{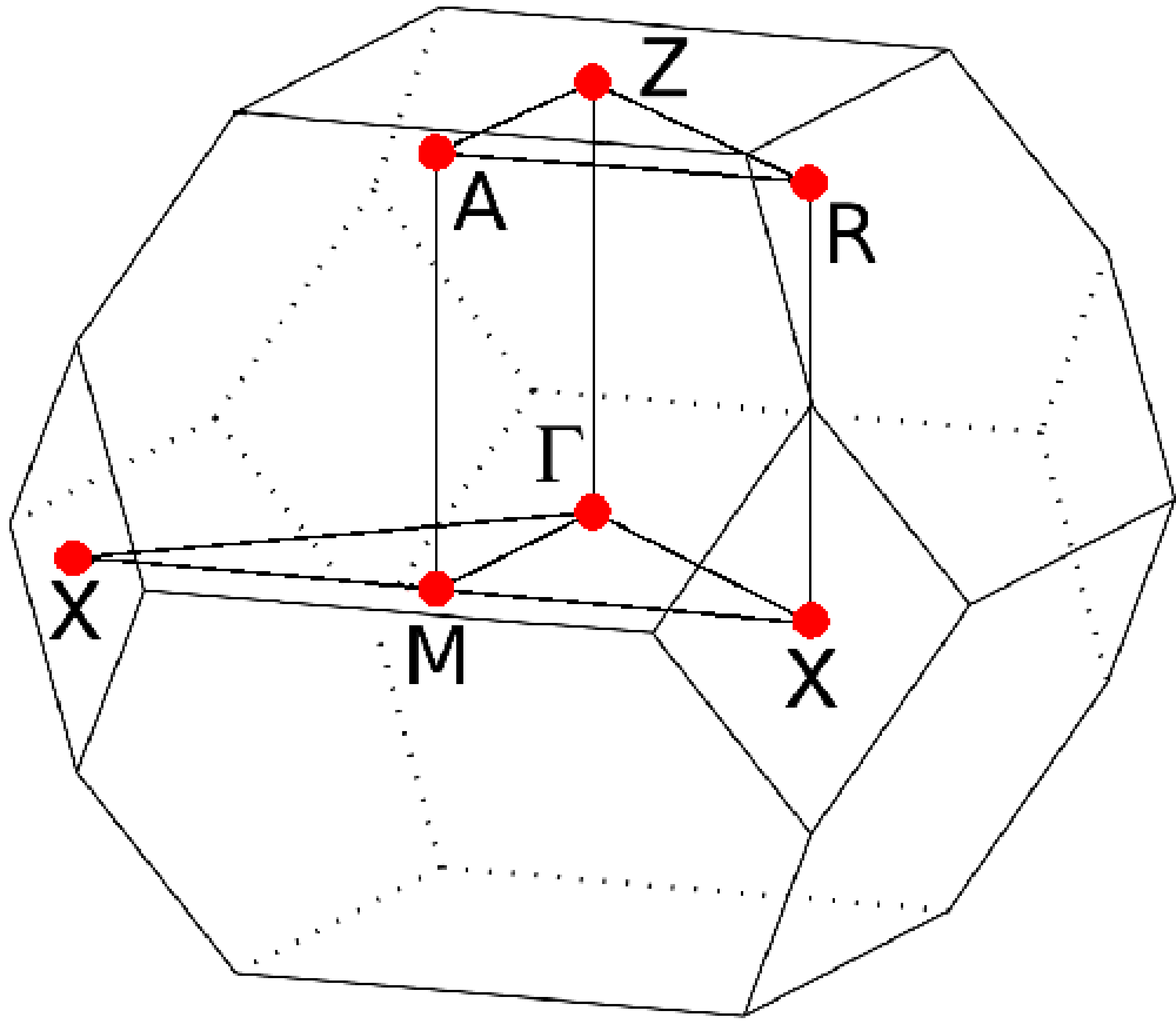}}
\end{picture}
\caption{(Color online) Brillouin zone of tetragonal \ce{BaFe2As2}, corresponding to Fig.~\ref{Fig_CPA}.}\label{Fig_BZTet}
\end{figure}

In Fig.~\ref{Fig_BZTet} the tetragonal Brillouin zone of \ce{BaFe2As2} is shown with its definition of high symmetric points corresponding to Fig.~\ref{Fig_CPA}.

In Table \ref{Tab_Lattice} all structure parameters used for the calculations are summarized for the sake of completeness. For the undoped compound only the experimental lattice values from Rotter \etal \cite{RTJ+08} were used. For increasing Co concentration the change of the $c$-axis was extrapolated on basis of the experimental results from Sefat \etal \cite{SJM+08}, while the change of the As position was extrapolated based on single crystals x-ray diffraction data of Merz \etal \cite{MSN+13}, accounting for the clinching effect of the $c$-axis. Similar trends of the lattice parameters are also found elsewhere in literature.\cite{NTY+08, KOK+09} Note that $a$ and $b$ were held constant over the whole doping regime.

\begin{table}[b]
\caption{Summary of all used structure parameters of orthorhombic \ce{Ba(Fe_{1-$x$}Co_$x$)_2As2} (space group 69, \textit{Fmmm}).}
	\vspace{0.5cm}
	\begin{tabular}{C{2.2cm}|C{3.2cm}|C{2.2cm}}
	{Cobalt ratio $x$}&{Lattice constants [\AA]}&{As position}\tabularnewline
	\hline	
	$x = 0.000 $ & $a = 5.6146$\\$b = 5.5742$\\$c=12.9453$ & $z = 0.3538$ \tabularnewline
	\hline
	$x = 0.025 $ & $c=12.9349$ & $z = 0.3536$ \tabularnewline
	$x = 0.050 $ & $c=12.9244$ & $z = 0.3534$ \tabularnewline
	$x = 0.075 $ & $c=12.9140$ & $z = 0.3531$ \tabularnewline
	$x = 0.100 $ & $c=12.9035$ & $z = 0.3529$ \tabularnewline
	$x = 0.125 $ & $c=12.8931$ & $z = 0.3527$ \tabularnewline
	$x = 0.150 $ & $c=12.8827$ & $z = 0.3525$ \tabularnewline
	$x = 0.175 $ & $c=12.8722$ & $z = 0.3522$ \tabularnewline
	$x = 0.200 $ & $c=12.8618$ & $z = 0.3520$ \tabularnewline
	$x = 0.213 $ & $c=12.8564$ & $z = 0.3519$ \tabularnewline
	$x = 0.225 $ & $c=12.8514$ & $z = 0.3518$ \tabularnewline
	$x = 0.250 $ & $c=12.8409$ & $z = 0.3516$ \tabularnewline
	\end{tabular}
\label{Tab_Lattice}
\end{table}

%
%
\newpage
\bibliographystyle{aipnum.bst}

\end{document}